\documentclass{optica-article}

\journal{opticajournal} % for journals or Optica Open

\articletype{Research Article}

\usepackage{upgreek}
\begin{document}

\title{Radiation hardness of open Fabry-P{\'e}rot microcavities}

\author{
Fernanda C. Rodrigues-Machado,\authormark{1,$\dagger$,*} 
Erika Janitz,\authormark{2,$\dagger$,*} 
Simon Bernard,\authormark{1}
Hamed Bekerat,\authormark{3,4}
Malcolm McEwen,\authormark{5}
James Renaud,\authormark{5}
Shirin A. Enger,\authormark{3}
Lilian Childress,\authormark{1}
and Jack C. Sankey\authormark{1}}

\address{
\authormark{1}Physics Department, McGill University, 3600 Rue University, Montréal, QC H3A 2T8, Canada\\
\authormark{2}Department of Electrical and Software Engineering, University of Calgary, 2500 University Drive NW, Calgary, AB T2N 1N4, Canada\\
\authormark{3}Medical Physics Unit, McGill University, 1001 Boul. Décarie,
Montréal, QC H4A 3J1, Canada\\
\authormark{4}Radiation Oncology Department, Jewish General Hospital, 3755 Chem. de la Côte-Sainte-Catherine, Montréal, QC H3T 1E2, Canada\\
\authormark{5}Metrology Research Centre, National Research Council of Canada, Ottawa, Ontario, Canada\\
\authormark{$\dagger$}Equal contributions
}

\email{\authormark{*}fernanda.rodriguesmachado@mail.mcgill.ca} 

% use {asbstract*} to suppress the copyright line. Copyright information will be added in production

\begin{abstract*} 
High-finesse microcavities offer a platform for compact, high-precision sensing by employing high-reflectivity, low-loss mirrors to create effective optical path lengths that are orders of magnitude larger than the device geometry. Here, we investigate the radiation hardness of Fabry-P{\'e}rot microcavities formed from dielectric mirrors deposited on the tips of optical fibers. The microcavities are irradiated under both conventional ($\sim$~ 0.1~Gy/s) and ultrahigh (FLASH, $\sim$~ 20~Gy/s) radiotherapy dose rates. Within our measurement sensitivity of $\sim$~40~ppm loss, we observe no degradation in the mirror absorption after irradiation with over 300~Gy accumulated dose. This result highlights the excellent radiation hardness of the dielectric mirrors forming the cavities, enabling new optics-based, real-time, \emph{in-vivo}, tissue-equivalent radiation dosimeters with $\sim$10 micron spatial resolution (our motivation), as well as other applications in high-radiation environments.
\end{abstract*}

%%%%%%%%%%%%%%%%%%%%%%%%%%  body  %%%%%%%%%%%%%%%%%%%%%%%%%%
\section{Introduction}

External beam radiotherapy (EBT) is one of the most common treatments for cancer worldwide. It relies on the selective destruction of tumoural cell DNA via irradiation~\cite{hendee2013radiation} with energetic~(MeV) photons, electrons, or protons~\cite{newhauser2015physics}.
For all applications, it is critical to accurately estimate -- ideally by real-time, \textit{in vivo} measurement~\cite{mijnheer2013} -- the radiation dose deposited in the tumour and surrounding tissue.

Many dosimeters have been explored in both literature and clinical practices \cite{hogstrom2006review,ashraf2020dosimetry}, but all come with important limitations that may either result in undesired irradiation of surrounding healthy tissues or imprecision in the delivered dose~\cite{holmberg2021}, or make them incompatible with new ultrahigh dose rate schemes (FLASH~\cite{favaudon2014,bourhis2019,vozenin2019}). These issues emerge from technical challenges~\cite{podgorsak2005,azangwe2014,daniel2022,devic2016,rampado2006,martivsikova2008,suchowerska2001} including (i)~dissimilarity between the sensing medium and tissue, which leads to dose rate and beam-energy dependencies, and the need for multiple calibrations, (ii)~sensor dimensions or materials that perturb beam fluence, limit spatial resolution, and/or preclude \emph{in-vivo} measurement, (iii)~directional dependencies, and (iv)~lack of real-time readout.

All four challenges could potentially be addressed using a fiber-coupled Fabry-P{\'e}rot optical microcavity~\cite{hunger2010} filled with water -- a tissue-equivalent medium -- probed on resonance to continuously monitor for the microseconds-long transients in absorption associated with hydrated electrons generated by passing radiation~\cite{stein1952,hart1970,hart1971,turi2012,svoboda2020, megroureche2023}. 
While hydrated electron spectroscopy has long been a known modality of radiation detection in water~\cite{hart1962,megroureche2023}, the long optical paths (${\sim 10-100}$~cm) required to obtain a measurable absorption signal have hampered practical application to dosimetry. By enclosing the water within a high-finesse optical microcavity, a long optical path can be realized with a sensor that is shorter by many orders of magnitude. In particular, fiber cavities, which employ mirrors fabricated on the tips of optical fibers, routinely achieve an optical path length enhancement of ${\sim 10,000}$ or more with active volumes as small as ${\sim (10~\upmu\text{m})^3}$ and external dimensions nominally limited by the~${\sim 125\ \upmu}$m diameter of standard single-mode optical fibers. 
This small sensor size, coupled with alignment-free fiber delivery of optical signals, lends itself naturally to endoscopic applications. 

Before considering such a detector, it is essential to verify that the cavity mirrors do not significantly degrade when exposed to relevant doses of radiation (i.e., that they exhibit ``radiation hardness''). To get a sense of scale for the required mirror performance, we note that folding the optimal $\sim$meter-scale water path \cite{megroureche2023} into a sub-millimeter optical cavity requires finesse of at least $\sim$3,000, corresponding to mirror losses below $\sim$1,000 ppm (note this scales with cavity length). If clinical radiation introduces additional losses at this level, this technology will be hampered by the need to recalibrate or replace mirrors frequently. 

In this paper, we verify the radiation hardness of finesse $>$8,000 Fabry-P{\'e}rot fiber cavities formed by dielectric micromirrors under extreme conditions. Specifically, we observe no measurable increase in cavity losses (within our $\sim40$~ppm sensitivity) across multiple devices and for accumulated doses up to ${\sim 307}$~Gy delivered under clinical and FLASH conditions. Radiation hardness is observed for both direct (electron beam) and indirect (photon beam) radiation. Beyond enabling a new paradigm for oncological dosimetry, these robust mirrors may also be used for quantum optics and sensing applications in other high-radiation environments including space~\cite{girard2018recent}, nuclear, and / or defense scenarios.

%%%%%%%%%%%%%%%%%%%%%%%%%%
\section{Experimental Setup and Methods}
\label{sec:methods}

\begin{figure}[ht!]
\centering\includegraphics[width=1\textwidth]{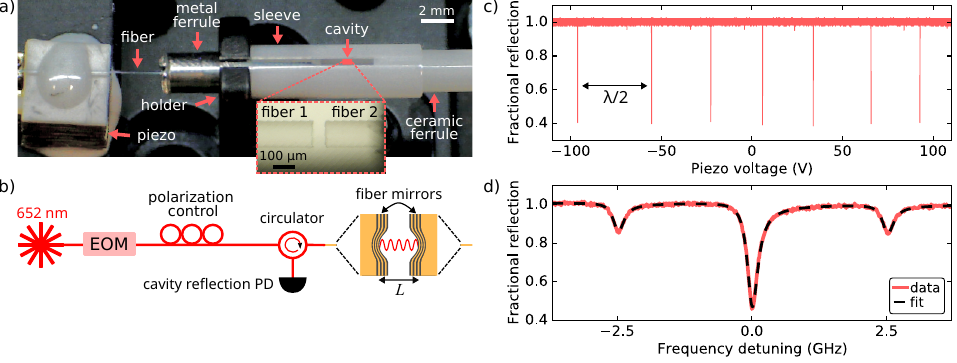}
\caption{(a)~Microcavity assembly (inset: close-up of the cavity fibers). 
(b)~Optical circuit. Cavity length $L$ is swept with a shear piezoelectric actuator in (a). Laser light (652 nm) is coupled to the cavity through one fiber mirror, while reflected light (50-100~$\upmu$W) from the same mirror is collected through an optical circulator by a photodiode (PD). An electro-optical modulator (EOM) generates GHz-frequency sidebands on the laser light, providing a frequency reference for cavity linewidth extraction in (d). 
(c)~Reflected power (normalized by the off-resonant value) as a function of piezo drive voltage, exhibiting 7 longitudinal cavity modes separated by ${\lambda/2=326}$~nm (i.e., one~FSR).  
(d)~High-resolution scan of reflected power (red data) versus detuning for a single mode in (c) with 2.5~GHz phase modulation applied to the laser. The black dashed line is a fit to three asymmetric Lorentzian profiles~\cite{janitz2015fabry}.}
\label{fig:setup}
\end{figure}

We test air-filled fiber-fiber cavities (FFCs) comprising two concave micromirrors fabricated on the tips of opposing single-mode pure silica optical fibers (Thorlabs 630HP; Fig.~\ref{fig:setup}a inset). Small ($\sim$50 $\upmu$m) radius-of-curvature dimples are created on cleaved fiber tips using a CO$_2$ laser ablation process~\cite{hunger2010}, and subsequently coated with dielectric Bragg mirrors having alternating layers of Ta$_2$O$_5$ and SiO$_2$~\cite{janitz2015fabry}.
The small dimples in this batch of fiber mirrors limit our stable cavity lengths to $<$30 $\upmu$m. However, since irradiation can only introduce losses via the coating materials (see below), our results apply to other mirror geometries and cavity lengths as well.
We probe the cavities with a 652~nm laser (Toptica DL Pro); 
this wavelength is chosen because it is near the maximum ratio of absorption from hydrated electrons~\cite{hart1971} and water~\cite{segelstein1981}. The Bragg stack is designed to transmit $\sim$60 ppm at this wavelength, corresponding to finesse $\sim$50,000. However, the combination of material absorption, scattering losses, and mirror shape provide a comparable loss \cite{benedikter2015transverse,janitz2015fabry}, reducing the finesse to $\sim$10,000-20,000 for this particular batch of cavities.

Passive cavity alignment is achieved using two ferrules mounted in a C-shaped sleeve, where one fiber is glued to a long-range shear piezoelectric actuator (Thorlabs PN5FC1; Fig.~\ref{fig:setup}a) used to tune the cavity length $L$ over ${\sim3\ \upmu}$m, corresponding to $\sim10$ free spectral ranges (FSRs). Laser light is injected through one fiber, and light reflected back into the same fiber is detected through an optical circulator (OZ Optics) with a photodiode (Thorlabs DET10A or Thorlabs PDA8A; Fig.~\ref{fig:setup}b). An example cavity reflection signal is shown in Fig.~\ref{fig:setup}c, exhibiting 7 narrow resonances over the piezo scanning range. A high-resolution scan of individual dips reveals an asymmetric line shape (Fig.~\ref{fig:setup}d) associated with imperfect alignment between the cavity and fiber modes~\cite{janitz2015fabry}.  

Mirror radiation hardness is tested by measuring the cavities' finesse $\mathcal{F}$ and linewidth $\Delta \nu$ before and after irradiation. These quantities depend on the power transmission $T_1$, $T_2$ through each mirror and the round-trip power loss $\delta$ (absorption and scattering) as ${\mathcal{F} \approx 2\pi/\mathcal{L}}$ and ${\Delta \nu \approx c\mathcal{L}/(4\pi L)}$, where $L$ is cavity length, and  $\mathcal{L} = T_1 + T_2 + \delta$ is the \textit{total} round-trip loss. 
Note these quantities do not vary with radiation-induced
changes in the surrounding silica fibers, and (for a fixed cavity alignment) any finesse or linewidth variation can be entirely attributed to a change in mirror loss. 

We extract $\mathcal{F}$ for a given cavity mode from the ratio of the spacing between adjacent resonances (i.e., one FSR; see Fig.~\ref{fig:setup}c) to the linewidth of the resonance (Fig.~\ref{fig:setup}d). 
To reduce vibrations, we drive the piezo with a continuous sinusoid, and calibrate the relative mirror position ($x$-axis) from the raw data (Fig.~\ref{fig:setup}c) by fitting the resonance locations -- known to occur every half wavelength (326 nm) -- to a sinusoid. 
To estimate~$\Delta \nu$, we add sidebands to the laser with an electro-optical modulator (EOM) at a single convenient frequency between ${2.5-6.5}$~GHz (2.5~GHz in Fig.~\ref{fig:setup}d), producing three peaks of known separation in the reflected signal, allowing us to calibrate the detuning between the cavity resonance and laser light.

%%%%%%%%%%%%%%%%%%%%%%%%%%
\section{Conventional dose rate exposure}\label{sec:conventional}

\begin{figure}[ht!]
\centering\includegraphics[width=1\textwidth]{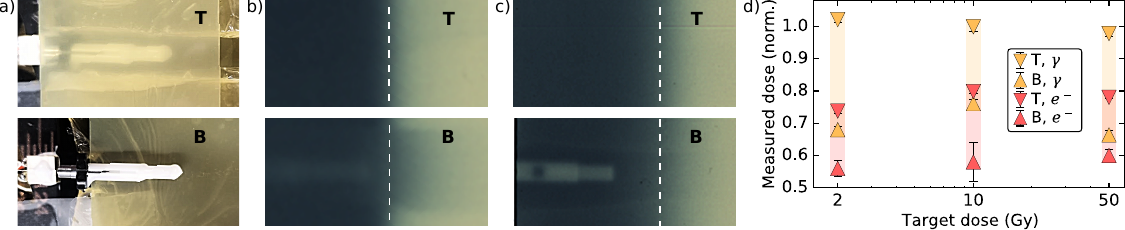}
\caption{Film dosimetry for conventional dose rate irradiation. (a)~Radiochromic film is placed on top of~(T) and below~(B) the cavity device using 0.5~cm Superflab spacers. (b)-(c)~Typical top~(T) and bottom~(B) films after exposure to 1000~MU (10~Gy) of (b)~gamma rays and (c)~electrons; the white dashed line represents the edge of the output field (the left, darker side is exposed).
(d)~Dose measured by the top~(T) and bottom~(B) films exposed to photon ($\gamma$; yellow) and electron ($e^-$; red) radiation, normalized to the target doses of 2, 10 and 50~Gy.
}
\label{fig:irradiation}
\end{figure}

\begin{figure*}[ht!]
\centering\includegraphics[width=1\textwidth]{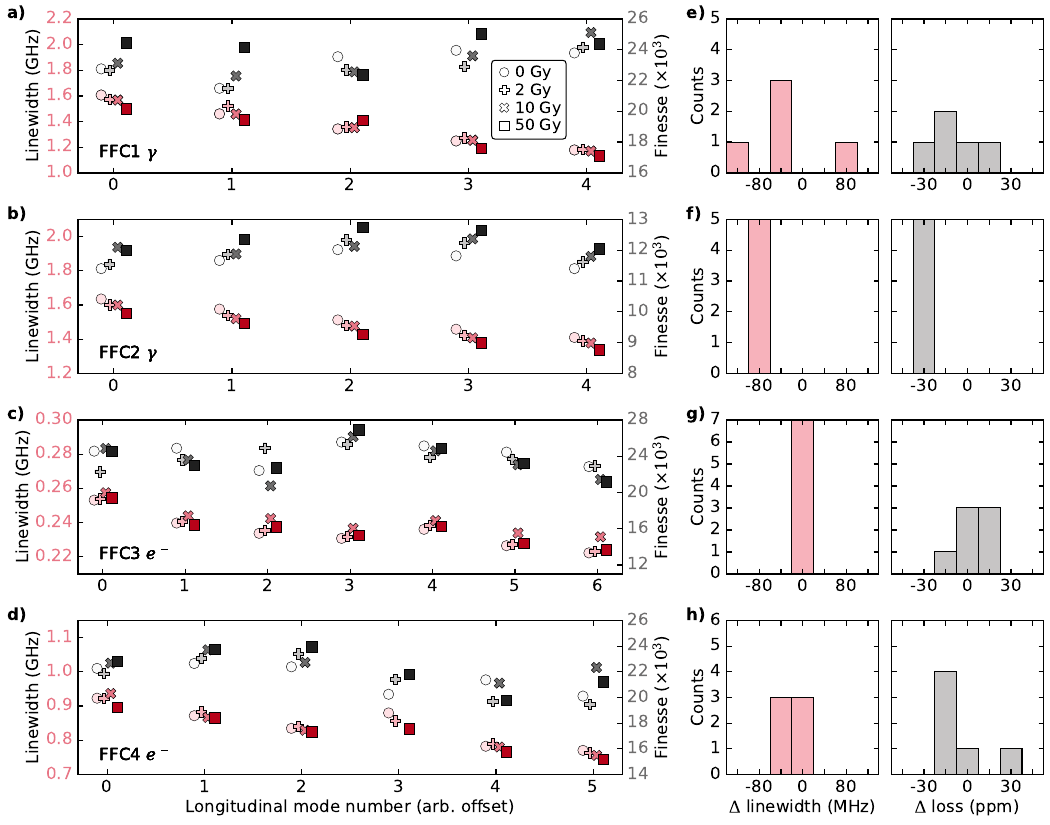}
\caption{
Changes in microcavity loss after irradiation at conventional dose rates. \mbox{(a)-(d)}~Linewidth (red; left axis) and finesse (grey; right axis) measurements for cavities exposed to (a)-(b)~photon ($\gamma$; 1000~MU/min) and (c)-(d)~electron ($e^-$; 600~MU/min) irradiation for total target doses of 0 (pre-irradiation), 2, 10 and 50~Gy. Color gradient and an artificial offset from left to right (to increase visibility) denotes increasing dose per exposure. 
Each point is an average of $\sim 50$ fitted scans. Error bars (standard error on the mean) are within marker size ($\lesssim 0.5\%$). RMS deviations between scans are~$\lesssim 3\%$, with individual fit errors~$\lesssim 1\%$.
(e)-(h) Change in linewidth (red) and round-trip loss ($2\pi/\mathcal{F}$; grey) accumulated between measurements before the first exposure (0~Gy) and after the final 50~Gy exposure (62~Gy in total) for FFC1-4, respectively.}
\label{fig:results_conv}
\end{figure*}

We first test the fiber mirrors under conventional radiotherapy dose rates with both photon and electron beams using a clinical linear particle accelerator (Varian Trilogy Clinac, Varian Medical Systems) with a radiation field size of ${3\times 5}$~cm$^2$ at a source-to-surface distance (SSD) of 1~m. 
The photon irradiation is set at 6~MV and dose rate of 1000~MU/min (1~MU~=~1~cGy of absorbed dose to water at the depth of maximum dose, for SSD of 1~m and field size of ${10\times 10}$~cm$^2$); the electron irradiation is set at~6 MeV, with a dose rate of 600 MU/min.
During irradiation, we surround the devices with a standard bolus material (Superflab), that mimics human tissue and serves as a build-up medium, to ensure the appropriate radiation dose is delivered to the cavity vicinity. 
We use a 1-cm-thick top layer for the photon beam and a 0.5-cm-thick top layer for the electron beam to account for differences in particle penetration depth~\cite{ziegler2010srim}. To test mirror hardness under conventional therapeutic conditions, the cavities are irradiated to 200, 1000 and 5000 MU, corresponding to target doses of $\sim$~2, 10, and 50~Gy. These values are chosen to span relevant clinical doses: 2~Gy corresponds to a typical partial dose delivered to patients in a single session, and 50~Gy is roughly the maximum clinical accumulated dose used in conventional radiotherapy schemes. 
 
The dose delivered to the cavities is estimated using EBT3 radiochromic film (Ashland). We determine bounds on this value using film below (B) and on top of (T) the devices~\cite{devic2005}, as shown in Fig.~\ref{fig:irradiation}a; the measurements have an uncertainty of 2\% according to the film dosimetry protocol. Example exposed films are shown in Fig.~\ref{fig:irradiation}b-c for 10~Gy photon and electron irradiation, respectively.
Due to the shorter penetration depth of electron irradiation, the device outline (C-clamp and ferrules) is more clearly visible on the bottom film for this modality. A summary of normalized delivered dose vs.~target dose output by the linear accelerator (LINAC) is shown in Fig.~\ref{fig:irradiation}d. 
The shaded regions between top and bottom films represent uncertainty in our calibration, corresponding to a maximum error of $41\%$ for electron therapy (at 10~Gy) and $42\%$ for photon therapy (at 2~Gy), ensuring that the delivered dose in the cavity vicinity is within a factor of two of the target value.

Figure~\ref{fig:results_conv}a-d shows results for the four devices (FFC1-4; cavity lengths $\approx$~5, 9, 27 and 8~$\upmu$m, respectively) before irradiation (0~Gy, light circles) and after each of the three tested doses (2, 10 and 50~Gy, from lighter to darker markers, respectively). Note that these labels indicate the target fractional dose of each irradiation, and the cumulative doses are 0, 2, 12, and 62~Gy, respectively. Devices 
FFC1-2 (Fig.~\ref{fig:results_conv}a-b) are exposed to gamma rays, while devices FFC3-4 (Fig.~\ref{fig:results_conv}c-d) are exposed to electrons. These quantities are measured for multiple adjacent (longitudinal) cavity modes, with each marker showing the average between $\sim 50$ acquisitions (statistical errors, i.e., the standard error on the mean, are within the marker size). The linewidth exhibits an overall downward trend with increasing cavity length (i.e., increasing mode number), as expected from the associated increase in round-trip time (which narrows resonances). Deviations from this trend (more visible in the nominally constant finesse) are attributed to mode-to-mode differences in alignment, transverse mode structure \cite{benedikter2015transverse}, and the resulting variations in the spatial distribution of the cavity mode on the (somewhat) non-uniform mirror coatings \cite{mader2015a}. 
Figure~\ref{fig:results_conv}e-h shows histograms of the \textit{total} accumulated changes in cavity linewidth (red) and round-trip loss ${2\pi/\mathcal{F}}$ (grey) between the non-irradiated cavities (0~Gy data) and last conventional dose rate exposure (50~Gy data, 62~Gy total) for these cavity modes in each device. 
Importantly, we observe no consistent correlation between radiation dose and cavity linewidth for either photon or electron beam therapies. We attribute these overall ${\lesssim 5\%}$ variations to a combination of drifts in transverse cavity alignment (this changes which regions of the mirror surfaces participate in determining the finesse) and imperfect coupling to a single polarization mode of the cavity (coupling is maximized with the use of fiber paddles before the devices). 
These drifts place a systematic uncertainty on our ability to observe added losses of $\lesssim 40$~ppm.

%%%%%%%%%%%%%%%%%%%%%%%%%%
\section{Ultrahigh dose rate exposure}\label{sec:ultrahigh}

\begin{figure*}[ht!]
\centering\includegraphics[width=1\textwidth]{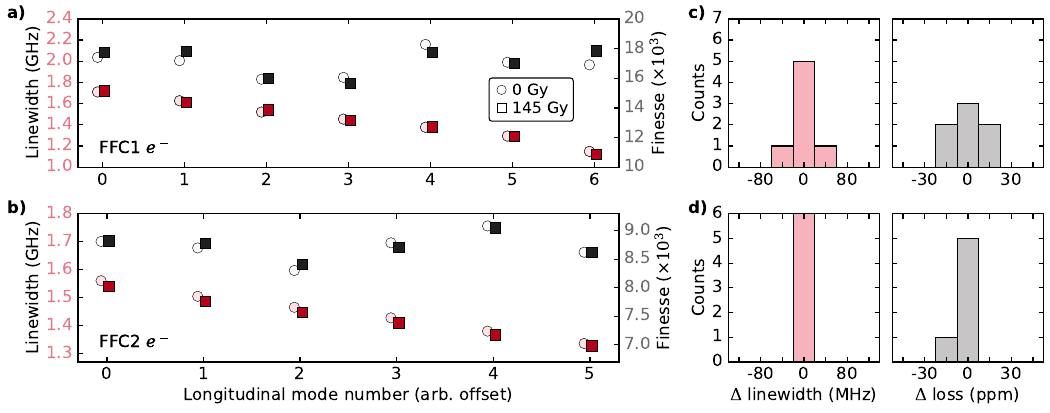}
\caption{Changes in microcavity loss after irradiation with ultrahigh dose rates. Cavity linewidth (red; left axis) and finesse (grey; right axis) as a function of longitudinal mode number for cavities FFC1 (a) and FFC2 (b). Light circles (dark squares) indicate pre (post) 145~Gy, 8~MeV electron radiation with an ultrahigh dose rate of ${21.4\pm0.6}$~Gy/s for~${6.8\pm0.1}$~s (${107 \pm 3}$~mGy/pulse at 200~Hz).  Markers are offset horizontally for improved visibility. 
Each point is an average of $\sim 50$ fitted scans. Error bars (standard error on the mean) are within marker size ($\lesssim 1\%$). RMS deviations between scans are~$\lesssim 7\%$, with individual fit errors~$\lesssim 1\%$.
(c)-(d)~Histograms representing the change in cavity linewidth (red) and round-trip loss ($2\pi/\mathcal{F}$; grey) between pre- and post-irradiation measurements.}
\label{fig:results_NRC}
\end{figure*}

The radiation hardness of the fiber mirrors is further tested at ultrahigh dose rates similar to those of FLASH. A modified clinical LINAC~\cite{Snyder2021Feb}  is used to irradiate FFC1 (now ${L\approx 6\; \upmu}$m) and FFC2 (now ${L\approx 12\; \upmu}$m) with 8~MeV electrons at ${21.4\pm0.6}$~Gy/s for~${6.8\pm0.1}$~s (${107 \pm 3}$~mGy/pulse at 200~Hz beam rate), resulting in a total absorbed dose of ${\sim145}$~Gy.
The delivered doses are estimated by a series of calorimetry measurements before and after irradiation, with a graphite calorimeter placed at the same source-to-field distance as the cavities. 
In this case, there is no need of a buildup medium since, for an 8~MeV electron beam, the dose absorbed on the surface of a water-equivalent medium is only $\sim1\%$ less than the dose measured at a depth of $\sim1$~mm (i.e., the approximate measurement depth of the calorimeter)~\cite{podgorsak2005}. 
The mass stopping power of silica is $\sim2\%$ greater than graphite at this beam energy~\cite{berger2017}, so we can assume that the dose absorbed by the fiber mirrors agrees with that measured by the calorimeter to within a few percent.

Figure~\ref{fig:results_NRC}a-b shows the cavities’ linewidth (red) and finesse (grey) before (light circles) and after (dark squares) this final ultrahigh dose rate irradiation. Markers show the average between fits of $\sim 50$ acquisitions for each longitudinal mode (error bars again within the marker size). Histograms on the right (Fig.~\ref{fig:results_NRC}c-d) show the changes in linewidth (red) and total round-trip cavity losses~$\mathcal{L}$ (grey) over all measured resonances. 
Our results are consistent with no measurable mirror degradation from ultrahigh dose rate exposure to within 20~ppm. Note the reduced drift here likely arises from the shorter (few days) time scale between pre- and post-irradiation measurements (the time scale for the 0-50 Gy sequence in Sec.~\ref{sec:conventional} was weeks).

These ultrahigh dose rate tests are performed on the same FFC1 and FFC2 devices as in Section~\ref{sec:conventional} after the clinical dose rate tests. In between these two tests, the cavities were additionally irradiated with ${\sim100}$~Gy at a rate of ${30\pm3}$~Gy/s. However, due to a mechanical malfunction, the cavity lengths changed such that the modes in Sections~\ref{sec:conventional} and \ref{sec:ultrahigh} are different, and so no conclusive data could be gleaned from this intermediate test. Nevertheless, we still observe high-finesse (up to nearly 20,000) after the cavities have been exposed to a total accumulated dose of 307~Gy.

%%%%%%%%%%%%%%%%%%%%%%%%%%
\section{Conclusion}
In summary, we have demonstrated the radiation hardness of an open Fabry-P{\'e}rot microcavity suitable
for future use in radiation oncology. The high-finesse dielectric mirrors studied here maintained low losses over a full range of clinically relevant irradiation conditions, including conventional therapies and FLASH-type exposure.
Limited by systematic drifts in our apparatus, we observed changes in loss below $\sim40$~ppm after 62~Gy delivered at conventional rates, and below $\sim20$~ppm after 145~Gy at FLASH-level rates. 

As mentioned, these drifts happen on a timescale of days to weeks, while the real-time absorption transients associated with hydrated electrons last microseconds. Due to the separation of time scales, the uncertainties reported here will not play a role in limiting the sensitivity of our envisioned dosimeter.

The compact footprint and high-sensitivity afforded by microcavity devices will open the door for high-precision, {\it in-situ}, tissue-equivalent, real-time dosimetry for radiation-based cancer treatments. Moreover, these mirror coatings should be suitable for sensing in other radiation-intensive environments, including space, nuclear, and/or defense scenarios.

%%%%%%%%%%%%%%%%%%%%%%%%%%
\begin{backmatter}

\bmsection{Funding} 
%This project was supported by funds from Fonds de recherche du Québec – Nature et Technologies (FRQNT, dossier \#2022-2023 - B2X - 319595), National Science and Engineering Research Council (NSERC, grant \# RGPIN 435554-2013, RGPIN-2020-04095), the Canada Research Chairs (grant \# 950-229003 and 950-231949),  Canada Foundation for Innovation (Innovation Fund 2015 Project \# 33488 and LOF/CRC 229003).

\bmsection{Acknowledgments} 
We thank Rigel Zifkin and Jiaxing Ma for fruitful discussions. 
FCRM acknowledges support from a Fonds de Recherche du Québec – Nature et Technologies (FRQNT) B2 scholarship (dossier \#2022-2023 - B2X - 319595). 
EJ acknowledges support from a Natural Sciences and Engineering Research Council of Canada (NSERC) postdoctoral fellowship (PDF-558200-2021).
LC acknowledges support from the National Science and Engineering Research Council of Canada (NSERC, grant \# RGPIN 435554-2013, RGPIN-2020-04095), the Canada Research Chairs (grant \# 950-229003 and 950-231949),  Canada Foundation for Innovation (Innovation Fund 2015 Project \# 33488 and LOF/CRC 229003).
JCS acknowledges support from the Natural Sciences and Engineering Research Council of Canada (NSERC RGPIN 2018-05635), Canada Research Chairs (CRC 235060), Canadian foundation for Innovation (CFI 228130, 36423), Institut Transdisciplinaire d'Information Quantique (INTRIQ), and the Centre for the Physics of Materials (CPM) at McGill.

\bmsection{Disclosures} 
The authors declare no conflicts of interest.

\bmsection{Data Availability Statement} 
The data presented here will be made available on the McGill Dataverse found at https://borealisdata.ca/dataverse/mcgill .

\end{backmatter}

%%%%%%%%%%%%%%%%%%%%%%% References %%%%%%%%%%%%%%%%%%%%%%%%%
\bibliography{biblio}

\end{document}